\documentclass[a4paper,fleqn,usenatbib]{mnras}

\usepackage[T1]{fontenc}
\usepackage{ae,aecompl}
\usepackage{graphicx}
\usepackage{amsmath}
\usepackage{amssymb}

\newcommand{\alphalow}{\alpha_\mathrm{low}}
\newcommand{\alphahigh}{\alpha_\mathrm{high}}
\newcommand{\mt}{m_\mathrm{t}}
\newcommand{\Msun}{\,\mathrm{M}_{\sun}}
\newcommand{\mratio}{m/\Msun}
\newcommand{\masyr}{\,\mathrm{mas\,yr^{-1}}}
\newcommand{\kms}{\,\mathrm{km\,s^{-1}}}
\newcommand{\pc}{\,\mathrm{pc}}
\newcommand{\Myr}{\,\mathrm{Myr}}
\newcommand{\percent}{per\,cent}
\newcommand{\Nbody}{$N$-body}

\newcommand{\secref}[1]{Section~\ref{#1}}
\newcommand{\figref}[1]{Fig.~\ref{#1}}
\newcommand{\tabref}[1]{Table~\ref{#1}}
\newcommand{\equref}[1]{equation~\eqref{#1}}

\defcitealias{Khalaj}{KB13}
\defcitealias{vanLeeuwen}{1}
\defcitealias{Madsen}{2}
\defcitealias{Netopil}{3}
\defcitealias{Barrado02}{4}

\title[Binary Fraction of Alpha Persei]{The Binary Fraction and Mass Segregation in Alpha Persei Open Cluster}

\author[Sheikhi et al.]{
Najmeh Sheikhi,$^{1}$\thanks{E-mail: najmesheikhi@yahoo.com}
Maryam Hasheminia,$^{1}$\thanks{E-mail: maryam.hashemi@alumni.znu.ac.ir}
Pouria Khalaj,$^{2}$\thanks{E-mail: pouria.khalaj@uqconnect.edu.au}
Hosein Haghi,$^{1}$ \thanks{E-mail: haghi@iasbs.ac.ir} 
\newauthor Akram Hasani Zonoozi,$^{1}$\thanks{E-mail: a.hasani@iasbs.ac.ir} 
and Holger Baumgardt$^{2}$\thanks{E-mail: h.baumgardt@uq.edu.au} 
\\
$^{1}$ Department of Physics, Institute for Advanced Studies in Basic Sciences (IASBS), PO Box 11365-9161, Zanjan, Iran\\  
$^{2}$ School of Mathematics and Physics, University of Queensland, Brisbane, QLD 4072, Australia\\
}

\begin{document}

\date{Accepted 2016 January 7.  Received 2015 December 19; in original form 2015 July 28}
\pagerange{\pageref{firstpage}--\pageref{lastpage}} \pubyear{2016}
\maketitle
\label{firstpage}

\begin{abstract}
We have obtained membership probabilities of stars within a field of radius $\sim3\degr$ around the centre of the open cluster Alpha Persei using proper motions and photometry from the \emph{PPMXL} and \emph{WISE} catalogues. We have identified 810 possible stellar members of Alpha Persei. We derived the global and radial present-day mass function (MF) of the cluster and found that they are well matched by two-stage power-law relations with different slopes at different radii. The global MF of Alpha Persei shows a turnover at $m=0.62\Msun$ with low and high-mass slopes of $\alphalow=0.50\pm0.09$ ($0.1<\mratio<0.62$) and $\alphahigh=2.32\pm0.14$ ($0.62\leq \mratio<4.68$) respectively. The high-mass slope of the cluster increases from $2.01$ inside $1\fdg10$ to $2.63$ outside $2\fdg2$, whereas the mean stellar mass decreases from $0.95$ to $0.57\Msun$ in the same regions, signifying clear evidence of mass segregation in the cluster. From an examination of the high-quality colour-magnitude data of the cluster and performing a series of Monte Carlo simulations we obtained a binary fraction of $f_{\rm bin}=34\pm12$ {\percent} for stars with $0.70<\mratio<4.68$. This is significantly larger than the observed binary fraction, indicating that this open cluster contains a large population of unresolved binaries. Finally, we corrected the mass-function slopes for the effect of unresolved binaries and found low- and high-mass slopes of $\alphalow=0.89\pm0.11$ and $\alphahigh=2.37\pm0.09$ and a total cluster mass of $352\Msun$.
\end{abstract}



\section{Introduction}
The distribution of stellar masses at birth, the initial mass function (IMF), that has been found to be invariant is one of the most important dynamical properties of stellar populations. The IMF is a very important ingredient in the understanding of a large number of basic astronomical phenomena such as the formation and evolution of the first stars and galaxies. Although the determination of stellar MF is important in many branches of astrophysics there is no direct observational determination of the MF. What is observed is the individual or integrated light of objects. Transformation of this observable quantity into the MF thus relies on theories of stellar evolution \citep{Chabrier}. Therefore, the MF provides an important test of the stellar evolution theories. 

\par In young stellar systems, most of the low-mass stars will still be in their pre-main-sequence evolutionary stage, and the MF has not yet affected by the stellar and dynamical evolution, so one may expect to derive the global stellar IMF from their present-day MF.

\par Since most Galactic open clusters are disrupted on a time scale of a few hundred Myrs, most present clusters tend to be relatively young. Stars in open clusters lie roughly at the same distance with the same age, chemical abundance and reddening, at least to a first approximation. Therefore these clusters are considered as excellent laboratories for our understanding of the star formation process, stellar evolution and the dynamics of stellar systems. Especially for nearby open clusters, the selection of member stars is easy owing to average large proper motions they have.

\par Another important distribution function characterizing the stellar populations is the distribution functions of semi-major axes, mass ratios and eccentricities in binary systems which seems to be quiet invariant to the physical conditions of star formation \citep{Kroupa11}. Observations and theoretical works indicate that stars in star clusters may be born with initially very large binary fractions and thus the majority of stars are found in binary or multiple systems \citep{Duquennoy, Kroupa95a, Kroupa95b, Griffin, Halbwachs, Kouwenhoven05, Goodwin, Rastegaev}. Indeed, many Milky Way open and globular clusters show a binary fraction with a rising binary frequency towards the cluster core, which is interpreted to be the result of mass segregation (e.g. \citealt{Mathieu, Geller, Milone}). Binary stars, either primordial or dynamically formed during close encounters between single stars, can affect the observational parameters of a star cluster, such as the velocity dispersion and the stellar MF. Therefore, characterization of the binary fraction in star clusters is of fundamental importance for many fields in astrophysics.   

\par For example, the presence of binaries affect the dynamical mass estimation of a star cluster. The dynamical mass of a star cluster can be determined from the virial theorem, using the measured half-mass radius and line-of-sight velocity dispersion of the cluster. This dynamical mass may be a significant overestimation of the cluster mass if the contribution of the binary orbital motion is not taken into account. This is because the stars in binary systems exhibit not only motion in the gravitational potential of the cluster, but also orbital motion in the binary system. Indeed, in a cluster consisting of single stars, the velocity dispersion traces the motion of each particle, (i.e. star) in the cluster potential, while in a cluster with binary stars, we do not measure the motion of each particle (i.e., the binary centre-of-mass), but of the individual binary components. These have an additional velocity component due to their orbital motion, which may result in a major dynamical mass overestimation depending on the properties of the binary population of cluster (e.g., \citealt{bosch, Fleck, Apai}). This effect is more important if the typical orbital velocity of a binary component is of order, or larger than, the velocity dispersion of the particles (single/binary) in the potential of the cluster. The results of simulations, indicate that the dynamical mass is overestimated by a factor of two or more for low-mass star clusters, while the dynamical masses of massive star clusters are only mildly affected by the presence of binaries \citep{Kouwenhoven08a, Kouwenhoven08b, Kouwenhoven09}.

\par Furthermore, the stellar MF of a star cluster is also affected by the presence of binaries. If the binarity is not taken into account in the determination of the  stellar masses, the combined system will be assigned a mass that is larger than the mass of the two single stars. This causes an unrealistic flattening in the MF slope \citep{Kroupa91, Kroupa93}. This is because an unresolved binary consisting of two main-sequence stars has a combined colour somewhere in between the colours of the two components, and a magnitude brighter than that of a single-star main sequence at this combined colour \citep{Kroupa92a}. The magnitude of this effect depends on the fraction of unresolved binaries and the mass distribution of the binary components.

\par The effect of unresolved binary and multiple systems on the MF slope is more evident in the low-mass range $m\leq0.5\Msun$ \citep{Kroupa92b, Kroupa93, Kroupa08}, while it is small in the high-mass end even for a binary fraction of $90$ {\percent} \citep{Zonoozi14, Khalaj}. A decrease of $\alpha$ at the low-mass end can be expected since many low-mass stars will be hidden in binaries with more massive companions.

\par High-quality colour-magnitude data for open clusters can be used to put constraints on the unresolved binary fraction. Since all stars in a cluster have the same metallicity and age, the intrinsic scatter in colour-magnitude data will be small and high-quality photometric data can reveal effects of binaries. A second sequence $0.75\,\mathrm{mag}$ brighter than the single-star main sequence is clearly observed in a number of Galactic clusters. Therefore, finding the data points near the second sequence, that are very likely to be unresolved binaries, is a useful tool for the identification of cluster binaries.

\par Determinations of the MF and binary fraction of star clusters of different ages allow us to verify the predictions from {\Nbody} simulations and is therefore very important for our understanding of the early evolution of binaries in star clusters.  Recently, \cite{Khalaj} (hereafter KB13) derived the global and radial MF of Praesepe open cluster and found clear evidence for mass segregation as the high-mass slope of the radial MF increases from the inner to outer radii in Praesepe. Moreover, from an inspection of the colour-magnitude diagram (CMD) and  performing a series of Monte Carlo simulations they found an overall binary fraction of $f_{\rm bin}=35\pm5$ {\percent} for this cluster. 

\par The present work is a follow-up paper to \citetalias{Khalaj}, using the same method to investigate the MF, mass segregation and binary fraction of another nearby open cluster Alpha Persei.  For this, we combine the astrometric and near and mid-infrared photometric data that have been described in \secref{sec:surveys}. We explain the methods for selecting candidate cluster members in \secref{sec:membership}. We derive the MF in \secref{sec:MF} and examine the effect of mass segregation in the cluster followed by the determination of the binary fraction in \secref{sec:binary}. A conclusion is given in \secref{sec:summary}.


\section{Previous studies}
\par Alpha Persei is a young nearby open cluster ($d=172.4\pc$; \citealt{vanLeeuwen}) with an age of $90\pm10\Myr$ \citep{Barrado02} and a tidal radius of $9.7\pc$ \citep{Makarov}. The cluster members have solar metallicity \citep{Netopil} and the interstellar reddening towards this cluster is estimated to be $E(B-V)=0.09\pm0.03$ \citep{Netopil}. It has been reasonably well studied, despite its small proper motion and a low galactic latitude ($b=-7\degr$). The first proper motion survey in the cluster was performed by \citet{Heckmann} down to $V\approx12$ and complemented by $UBV$ photometry from \citet{Mitchell}. Lower mass members were identified by \citet{Stauffer85} and \citet{Stauffer89} on the basis of their proper motions, $UBVRI$ photometry and radial velocities. \citet{Prosser92} investigated the Palomar photographic plates to extract new low mass members down to $V=18.8$. \citet{Prosser94} obtained additional lower mass candidates from a deeper optical survey that were assessed later using near-infrared photometry and spectroscopy \citep{Basri}. The cluster has been also studied in X-rays with the ROSAT satellite \citep{Randich,Prosser96,Prosser98a,Prosser98b}.

\par \citet{Barrado02} identified a list of possible low-mass members down to $0.035\Msun$ ($I_{c}\sim22$) for Alpha Persei based on optical and near infrared photometry. They found that the MF can be fitted by a power-law, i.e. $N(m)\sim m^{-\alpha}$, with an index of $\alpha=0.59\pm0.05$. \citet{Deacon} identified 302 new candidate members down to about $R=18$ using photometry and accurate astrometry from the SuperCOSMOS microdensitometer. Based on the high probability members, a power-law index of 0.86 was derived over the $0.15-1.0\Msun$ range. \citet{Makarov} selected 139 possible members using the astrometric data and photometry from the Tycho-2 and ground-based catalogues. He found that only $\sim20$ {\percent} of members with $K_{s}<11$ are binaries. In the magnitude range of $9.5<V<13.85$, a minimum spectroscopic binary fraction of $32$ {\percent} was detected among the sample selected from \cite{Heckmann} by \citet{Mermilliod}. Finally, \citet{Lodieu12b} selected 728 cluster candidates with membership probabilities higher than $40$ {\percent} down to $J=19.1$ based on the photometry and proper motions from the UKIDSS GCS-DR9. They fitted a log-normal MF \citep{Chabrier} with a goodness-of-fit $\chi^{2}=2.275$, characteristic mass of $0.34\Msun$ and a dispersion of $0.46$ in the $0.04-0.5\Msun$ range.

 
\section{The surveys}
\label{sec:surveys}
The present study is based on the proper motion and $JHK_{s}$ photometric data extracted from the PPMXL catalogue \citep{Roser} combined with $W1$ and $W2$ magnitudes of the Wide-field Infrared Survey Explorer (WISE) \citep{Wright}. 

\par PPMXL contains mean positions and proper motions on the International Celestial Reference System (ICRS) by combining the USNO-B1.0 catalogue \citep{Monet} and 2MASS all-sky catalogue of point sources \citep{Cutri} astrometry. The USNO-B1.0 is a catalogue that presents positions, proper motions and magnitudes in various optical passbands for 1,045,913,669 objects obtained during the last 50 years. The catalogue is expected to be complete down to $V=21$ \citep{Monet}. The Two Micron All-Sky (2MASS) is a near infrared survey that uniformly scans the entire sky in three wavebands at $J(1.25\mu m)$, $H(1.65\mu m) $, and $K_{s}(2.17\mu m)$. The observations for the survey were taken between 1997 and 2001. 2MASS photometric errors typically attain $0.10\,\mathrm{mag}$ at $J=16.2$ and $H=15.0$ (see \citealt{Soares}). PPMXL contains about 900 million stars and galaxies, some 410 million with 2MASS photometry, from the brightest stars down to about $V=20$. The typical individual mean errors of the proper motions range from $4\masyr$ to more than $10\masyr$ depending on observational history. 

\par WISE is a NASA mission that has scanned the sky at $3.4\mu m(W1)$, $4.6\mu m(W2)$, $12\mu m(W3)$ and $22\mu m(W4)$ mid-infrared passbands in 2010. WISE contains accurate photometry and astrometry for over 300 million objects. The dataset obtained by WISE constitutes an excellent resource for finding new brown dwarfs. The $W1$ and $W2$ WISE bands have been specifically designed to identify T dwarfs from background sources. USNO-B1.0, 2MASS and WISE are full-sky surveys which have been shown to be an ideal instrument to study star clusters.


\section{Membership Determination}\label{sec:membership}
We determine the stellar members of Alpha Persei using the proper motions and photometry from PPMXL catalogue and WISE survey. Alpha Persei has a tidal radius of about $9.7\pc$ \citep{Makarov} which corresponds to $\sim3.3$ degrees at the distance of Alpha Persei ($172.4\pc$; \citealt{vanLeeuwen}). Therefore, we limit our selection area to a radius of $3\fdg3$ around the centre of Alpha Persei. 
Our detection limit for each filter is $J=15.5$, $K_s=15.4$, $W1=14.5$ and $W2=14.6$, where the typical proper motion errors are $\sim4.5\masyr$ and $\sim8\masyr$ for bright and faint stars respectively. A comparison with \citet{Lodieu12a, Lodieu12b} shows that, at these magnitude limits, PPMXL is about $90$ {\percent} complete. $J=15.5$ corresponds to $m=0.1\Msun$, which is our lower-mass limit in this study.

\begin{figure}
    \centering
    \includegraphics[width=0.45\textwidth]{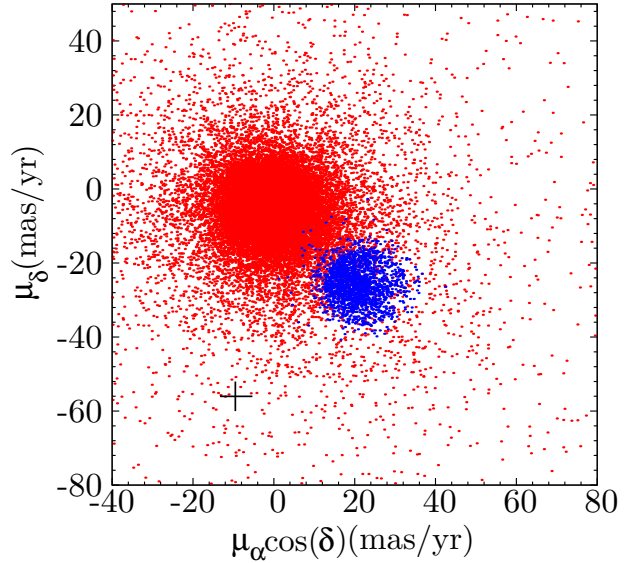} 
    \caption{The vector point diagram of proper motions for all stars in the field of Alpha Persei (red dots) and PM selected stars (blue dots) using \equref{eq:chi2}. A typical error bar of proper motions in PPMXL is shown by the black cross at the lower left corner of the plot.}
    \label{fig:PM}
\end{figure}

\begin{figure}
    \centering
    \includegraphics[width=0.40\textwidth]{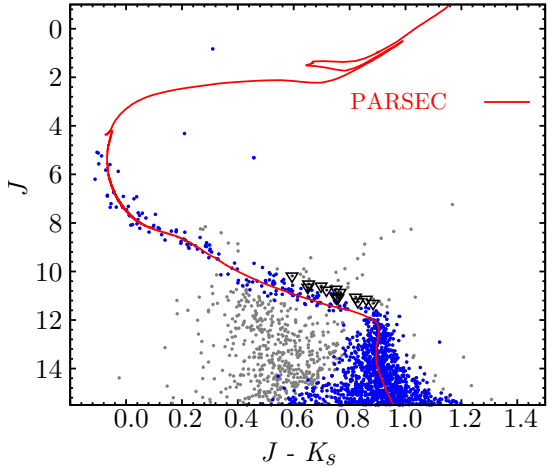}	
    \caption{The CMD of stars that satisfy our astrometric and photometric tests. The red line shows the best-fitting PARSEC isochrone. Photometrically selected stars, which are denoted by the blue dots, are located within a distance of $2.5\sigma$ from the isochrone. The black triangles are binaries (see \secref{sec:binary}).}
    \label{fig:CMD1}
\end{figure}

\begin{figure}	
	\centering
	\includegraphics[width=0.40\textwidth]{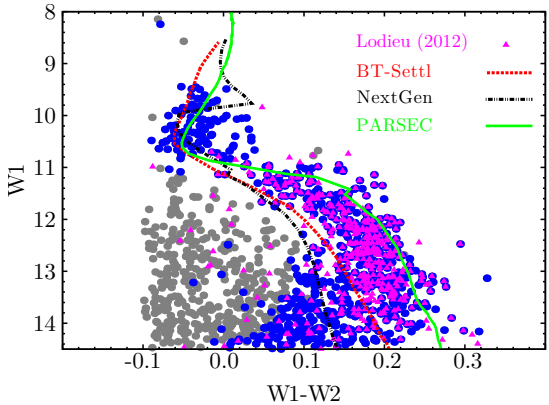}
	\caption{The CMD of low-mass stars of the cluster according to the data of WISE survey, which is obtained from a cross-matching between PPMXL and WISE data. The blue dots show stars that lie within $2.5\sigma$ of the PARSEC isochrone (green solid line). Grey dots correspond to field stars which do not belong to Alpha Persei as they fail our second photometric test. Magenta filled triangles show candidate stars of the cluster as found by \citet{Lodieu12b}. NEXTGen and BT-Settl isochrones are also shown for comparison. One can see that the PARSEC isochrone better describes the actual position of the possible stellar members of the cluster.}
	\label{fig:CMD2}
\end{figure}


\subsection{Astrometric selection}\label{sec:astrometric} 

\begin{table}
	\centering
 	\caption{The values that we have adopted in our astrometric and photometric tests for membership determination in Alpha Persei. ($\mu_{\alpha,\rm c}$, $\mu_{\delta,\rm c}$): average cluster proper motion; $(m-M)_0$: distance modulus; $A(V)$: extinction; $Z$: metallicity. 1: \citet{vanLeeuwen}; 2: \citet{Madsen}; 3: \citet{Netopil}; 4: \citet{Barrado02};}
	\label{tab:parameters}
	\begin{tabular}{ccc}
		\hline
		Parameter & Value & Ref. \\
		\hline		
		$\mu_{\alpha,\rm c}$ & $22.73\pm0.17\masyr$ &  \citetalias{vanLeeuwen} \\
		$\mu_{\delta,\rm c}$ & $-26.51\pm0.17\masyr$ &  \citetalias{vanLeeuwen} \\
		1D velocity dispersion & $0.7\pm0.13\kms$ & \citetalias{Madsen} \\		
		$(m-M)_0$ & 6.15 & \citetalias{vanLeeuwen} \\
		$A(V)$ & 0.28 & \citetalias{Netopil} \\
		$Z$ & 0.021  & \citetalias{Netopil} \\ 
		age & $100\Myr$ & \citetalias{Barrado02} \\
		\hline
	\end{tabular}	
\end{table}

\par The internal velocity dispersions of nearby clusters (less than $200\pc$) are much lower than those of field stars. As a result, one can use the velocity dispersion as a reliable tool to efficiently separate the possible stellar members of such clusters from the background stars. 

\par We use a chi-square test, similar to \citetalias{Khalaj}, to astrometrically filter out the background stars as follows:
\begin{equation}\label{eq:chi2}
	\chi^{2}= \frac{(\mu_{\alpha, \rm s}-{\mu_{\alpha, \rm c}})^{2}}{e^{2}_{\alpha,\rm s}+e^{2}_{\alpha,\rm c}+\sigma^{2}_{\alpha,\rm c}} + \frac{(\mu_{\delta,\rm s}-{\mu_{\delta, \rm c}})^{2}}{e^{2}_{\delta, \rm s}+e^{2}_{\delta, \rm c}+\sigma^{2}_{\delta, \rm c}}
	< 6.17
\end{equation}
where ($\mu_{\alpha,\rm s}$, $\mu_{\delta,\rm s}$) and ($\mu_{\alpha,\rm c}$, $\mu_{\delta, \rm c}$) correspond to the proper motion of each star (first pair) and the average cluster proper motion (second pair) along the right ascension ($\alpha$) and declination ($\delta$) axes. The corresponding errors of these parameters are denoted by ($e_{\alpha,\rm s}$, $e_{\delta,\rm s}$) and ($e_{\alpha,\rm c}$, $e_{\delta, \rm c}$) and the internal velocity dispersion of the cluster is ($\sigma_{\alpha,\rm c}$, $\sigma_{\delta,c}$). \tabref{tab:parameters} lists the values that we have adopted for ($\mu_{\alpha,\rm c}$, $\mu_{\delta,\rm c}$) and ($\sigma_{\alpha,\rm c}$, $\sigma_{\delta,\rm c}$) for membership determination in Alpha Persei. The value of $6.17$ corresponds to a confidence level of 95.4 {\percent} for a chi-distribution with two degrees of freedom.

\par In our astrometric test, we ignored stars whose proper motion errors were larger than $10\masyr$ (360 stars). We found 1934 stars in the field of Alpha Persei that satisfied our chi-square criterion given by \equref{eq:chi2}. Hereafter we refer to these stars as proper motion (PM) selected stars for convenience. Figure~\ref{fig:PM} shows these stars in a vector point diagram. One can see that our first membership criterion imposed by \equref{eq:chi2} has well detached the likely cluster members from the rest of the stars in the same field.

\subsection{Photometric selection}\label{sec:photometric}
In this section we perform a photometric test on our PM selected stars, using the latest version of the synthetic PARSEC isochrones{\footnote{version 1.2S}} \citep{Bressan, Tang14, Chen14} and the CMD 2.7 web interface{\footnote{available at \url{http://stev.oapd.inaf.it/cmd}}, assuming the following parameters for the cluster: age: $100\Myr$ \citep{Barrado02}; distance modulus $(m-M)_0$: $6.15$ \citep{vanLeeuwen}; extinction $A(V)$: $0.28$ \citep{Netopil}; metallicity $Z$: $0.021$. These parameters are summarized in \tabref{tab:parameters}.

\par In the CMD, we designate any PM selected star that resides within a distance of $2.5\sigma$ from the isochrone (red line in \figref{fig:CMD1}) as a possible member of Alpha Persei, where $\sigma$ is the photometric error for each star. In total, there are 1361 PM selected stars that pass our photometric test in Alpha Persei. These stars are denoted by blue dots in \figref{fig:CMD1}. The PARSEC isochrone describes the position of bright main sequence ($J<12$) stars in the CMD very well.

\par A possible shortcoming of such a photometric selection could be contamination by non-members in the low-mass range which is a direct consequence of the inherent uncertainties of the theoretical isochrones and the large photometric errors of the stars in this mass-range. To overcome this issue and further improve our selection, we used a second photometric test. For this test, we did a cross-match between the data of the WISE survey and the list of final members in the low mass range ($J>9.85$). We found that 1195 of our candidate members in the low-mass range had a counterpart in the WISE survey. In our second photometric test we use $W1$ and $W2$ magnitudes of WISE and applied a $2.5\sigma$ threshold from the PARSEC isochrone to filter out non-members. For low-mass stars, one can also use other isochrones such as NEXTGen \citep{Hauschildt} or BT-Settl \citep{Allard}. However, a comparison of the location of the previously known members from \citet{Lodieu12b} in a CMD, with the prediction of the theoretical isochrones showed that the PARSEC isochrone matches the actual position of the candidate stars better than NEXTGen or BT-Settl isochrones. Our final list of members contains 810 stars, shown by red dots in \figref{fig:CMD2}. \tabref{tab:members} shows the list of the possible stellar members. 

For comparison, after the astrometric test, we recover $\sim96$ {\percent} of all the stellar members found by \citet{Deacon} and $\sim66$ {\percent} of those found by \citet{Lodieu12b}, in a $3.3$-degree field around Alpha Persei and for $J>15.5$. After applying the second photometric test, these values slightly decrease to $\sim85$ {\percent} and $63$ {\percent} respectively. 

\begin{figure}	
	\centering
	\includegraphics[width=0.5\textwidth]{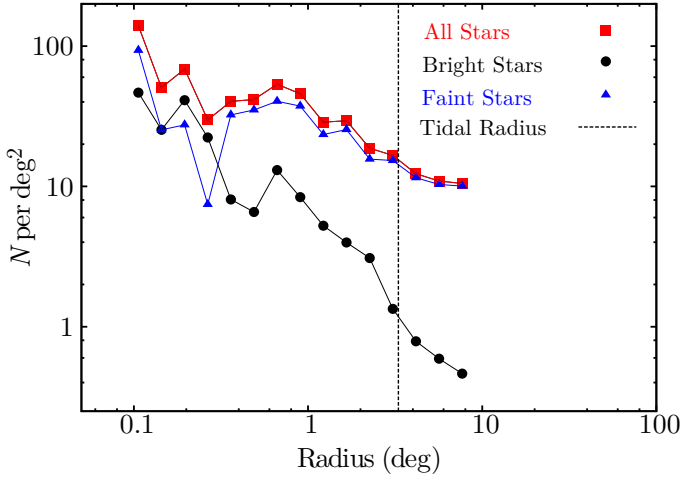}	
	\caption{The surface density of Alpha Persei as a function of cluster radius. In the last bin of radius, the surface density for all stars is 10.02 $\deg^{-2}$. The vertical dashed lines show the tidal radius of the cluster. The bright stars $(J\leq11)$ have masses $m\geq0.84\Msun$.}
	\label{fig:Density}
\end{figure}

\begin{table*}
    \caption{Celestial coordinates, proper motions, magnitudes and the masses of the possible members of Alpha Persei as found by this study, sorted by right ascension from PPMXL. Stellar masses are derived from the $J$ magnitudes using a PARSEC isochrone. In the last column, 'D' and 'L' letters correspond to \citet{Deacon} and \citet{Lodieu12b} respectively and show whether the members have also been designated by these studies. This table is available in its  entirety in the electronic version of the journal.}
    \label{tab:members}
    \centering
    \begin{tabular}{cccccccc}
        \hline
        RA(J2000) & Dec.(J2000) & $J$ & $K_{\rm s}$ & $\mu_{\alpha}\cos{\delta}$ & $\mu_{\delta}$ & Mass & found previously \\
        ($\deg$) & ($\deg$) & (mag) & (mag) & ($\masyr$) & ($\masyr$) & ($\Msun$) & \\
        \hline
        ... & ... & ... & ... & ... & ... & ... \\
        $48.863998$ & $+49.674999$ & $14.131\pm0.029$ & $13.204\pm0.029$ & $+24.4\pm4.1$ & $-29.9\pm4.1$ & $0.243$ & L \\
        $48.865002$ & $+48.810001$ & $15.424\pm0.088$ & $14.852\pm0.138$ & $+28.2\pm5.6$ & $-24.9\pm5.6$ & $0.098$ & - \\
        $48.867001$ & $+47.568001$ & $13.243\pm0.024$ & $12.360\pm0.023$ & $+16.8\pm3.9$ & $-24.2\pm3.9$ & $0.410$ & DL \\
        $48.900002$ & $+48.271999$ & $15.467\pm0.065$ & $14.673\pm0.088$ & $+16.4\pm5.5$ & $-24.9\pm5.5$ & $0.098$ & - \\
        ... & ... & ... & ... & ... & ... & ... \\
        $54.625999$ & $+48.320000$ & $13.037\pm0.023$ & $12.156\pm0.022$ & $+21.4\pm4.1$ & $-28.7\pm4.1$ & $0.450$ & DL \\
        $54.646259$ & $+48.593483$ & $08.104\pm0.020$ & $08.005\pm0.027$ & $+20.8\pm0.9$ & $-27.0\pm0.8$ & $1.818$ & - \\
        $54.650002$ & $+47.094002$ & $12.192\pm0.021$ & $11.278\pm0.018$ & $+17.8\pm4.1$ & $-25.0\pm4.1$ & $0.583$ & L \\
        $54.705002$ & $+48.145000$ & $12.719\pm0.023$ & $11.790\pm0.020$ & $+13.8\pm4.1$ & $-30.3\pm4.1$ & $0.507$ & L \\
        ... & ... & ... & ... & ... & ... & ... \\
        \hline

    \end{tabular}
\end{table*}

\begin{figure}  
    \centering
    \includegraphics[width=0.45\textwidth]{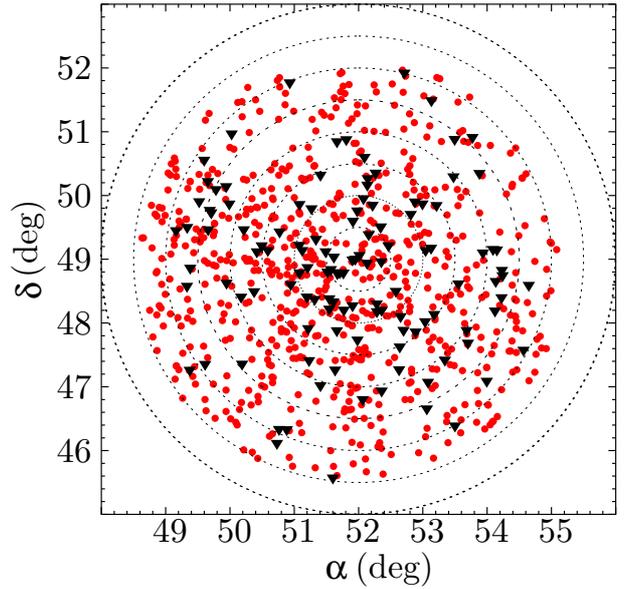} 
    \caption{The spatial distribution of possible members of Alpha Persei as found by our study. Red dots and black triangles correspond to low-mass and high-mass stars respectively. In agreement with \figref{fig:Density}, it can be seen that the stars show a concentration in the central region of the cluster. This concentration is more evident for bright (high-mass) stars.}
    \label{fig:Pos}
\end{figure}


\subsection{Contamination}\label{sec:contamination}
Using the membership selection criteria explained in sections \ref{sec:astrometric} and \ref{sec:photometric}, our final list of members should include most of the true cluster members. However, there might still be a number of field stars which we have erroneously been classified as members and one needs to account for such stars before deriving the underlying MF, total mass and other properties of the cluster. This is especially important for low-mass (faint) stars for which we suspect the contamination by field stars is larger than for bright stars. 

\par To estimate the number of field stars, we use the surface density distribution of Alpha Persei beyond its tidal radius. We first determine the cluster centre. The cluster centre can be defined as the region of the highest surface brightness, or the region containing the largest number of stars. We define the cluster centre as the density-weighted average of stellar positions ($x_i$) given by the following formula from \citet{vonHoerner}:
\begin{equation}
    \bar{x}_{d,j} = \frac{\sum_{i=1}^{N} x_i\rho_i^{(j)}}{\sum_{i=1}^{N} \rho_i^{(j)}}
\end{equation}
where $\rho_i^{(j)}$ is the local density estimator of order $j$ around star $i$. In our case $j$ is equal to 10 and we use the unbiased local density estimator of \citet{Casertano}. Using this method, the location of density centre is: \\ \\
$\alpha_{\rm c}=3^{\rm h}\ 27^{\rm m}\ 29^{\rm s} $, $\delta_{\rm c}=+48\degr\ 50\arcmin\ 24\arcsec$ \\
\par \figref{fig:Density} depicts the surface density of the Alpha Persei in a radius of $9\degr$ from the density centre of the cluster. We have used 15 radial bins to derive the surface density profile. Using the surface density of the last bin, which is well beyond the tidal radius of the cluster, we estimate the surface density of field stars. We found a density of $10.02\deg^{-2}$ for background contamination for stars with $J>15.5$. The background contamination for bright stars ($J<11$) is much lower ($\sim0.4\deg^{-2}$).

Assuming a constant surface density for field stars, one can simply multiply this value by the whole area within the tidal radius of the cluster, to get an estimate of contamination by field stars.

\par Using this method, we expect that out of 810 candidates found by our astrometric and photometric tests for Alpha Persei, about 469 of them to be contaminants, corresponding to a contamination level of $\sim58$ {\percent}. For comparison, in Alpha Persei, \citet{Lodieu05} reported a contamination level of about 70-80 {\percent} in an infrared-selected sample of stars and 28-40 {\percent} for in an optical-selected sample of stars from the KPNO/MOSA survey. 


\section{Mass function and Mass Segregation}\label{sec:MF}
Having found the number of possible contaminants in the field of Alpha Persei (\secref{sec:contamination}), we derive the MF of all the stars beyond the tidal radius of the cluster and normalize it to the total number of contaminants, and subtract it from the MF. We obtain an MF which has been corrected for contamination. \figref{fig:MF} illustrates the stellar MF which is corrected for contamination. As it can been from \figref{fig:MF}, Alpha Persei shows shows a turnover in its MF, indicating that low-mass and high-mass stars have a different MF. Over the whole mass range, the corrected MF of Alpha Persei, denoted by $\xi (m)$, can be fitted by a two-stage power-law function
\begin{equation}\label{eq:powerlaw}
  \xi (m)=\frac{dN}{dm}\propto
  	\begin{cases}
    	m^{-\alphalow} & $$m<\mt,$$\\
    	m^{-\alphahigh} & $$m\geq\mt$$
  	\end{cases}
\end{equation}
where $dN$ is the number of stars in interval $m$ and $m+dm$, $\alphalow$ and $\alphahigh$ represent the low-mass and the high-mass slopes{\footnote{$\alpha$ is the index of the power-law MF. On a log-log scale, it becomes the slope of the MF, hence we refer to $\alpha$ as slope.}} of the MF and $\mt$ is the turnover.

\begin{figure}
	\includegraphics[width=0.60\textwidth]{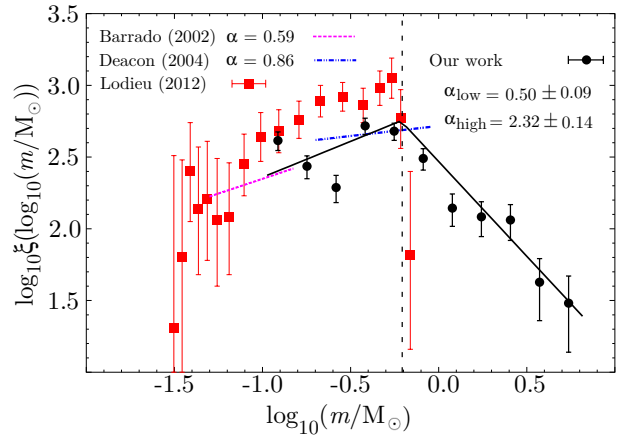}	
	\caption{The MF of Alpha Persei after correction for contaminants as found by our study as well as other studies. The vertical dashed line marks the MF turnover, i.e. $\mt=0.62\Msun$. The MF of low- and high-mass stars have power-law slopes of $0.50\pm0.09$ and $2.32\pm0.14$ respectively.}
    \label{fig:MF}
\end{figure}

\par We calculate $\alphalow$ and $\alphahigh$, as well as their errors and the location of turnover ($\mt$), using a method adopted by \citetalias{Khalaj} for Praesepe. This method is based on a maximum likelihood estimation combined with a Kolmogorov-Smirnov (K-S) test to obtain the goodness of fit and gives results which are more accurate than a least-squares method \citep{Clauset}. In addition, this method does not require to bin the data. We use formulae A1 and A3 in the Appendix of \citetalias{Khalaj} with a K-S test significance level of $5$ {\percent}. We only consider the main sequence stars to find the MF. The turnoff point in the CMD of Alpha Persei (\figref{fig:CMD1}) corresponds to a stellar mass of $4.68\Msun$.

\par Using this method we find that the MF of Alpha Persei is well described by a two-stage distribution over the $0.1-0.62\Msun$ and $0.62-4.68\Msun$ mass range with $\alphalow=0.50\pm0.09$ and $\alphahigh=2.32\pm0.14$. The high-mass slope of our fit is consistent with a Salpeter MF slope of $\alpha=2.35$. In the low-mass range, our result is very similar to $\alpha=0.59\pm0.05$ reported by \citet{Barrado02}, and is consistent with $\alpha=0.86^{+0.14}_{-0.19}$ derived by \citet{Deacon} within $1.5$ sigma. A comparison of our MF slopes and the results of \citet{Lodieu12b} is given in \secref{sec:simulation}. For comparison with \citet{Lodieu12b} we first fit a single power-law to the data points given in \figref{fig:MF}, since they do not give any MF slope in the paper. We find that the MF derived by \citet{Lodieu12b} has a MF slope of $\alpha=0.64\pm0.08$ in the mass-range of $0.1<\mratio<0.62$, which is in agreement with our $\alphalow$. The value of $\alpha=0.64\pm0.08$ is not reported in \citet{Lodieu12b} and is calculated by fitting a single power-law function to the data of \citet{Lodieu12b}.

\par Dynamical evolution and mass segregation can have a significant effect on the shape of the present day MF of open clusters. Equipartition leads to mass-loss with the preferential evaporation of low-mass cluster members. In order to examine the effect of the mass segregation in clusters one can find the MF for a sample of stars located at different radii with respect to the centre of the cluster and compare with the global one. For clusters with no mass segregation, one does not expect to see any change in the high-mass slopes of the MF, whereas in clusters with mass segregation the (low-mass) high-mass slope becomes (flatter) steeper at larger radii. In addition, mass segregated clusters have a lower mean stellar mass at larger radii (e.g. \citetalias{Khalaj} for Praesepe).

\par As shown in \figref{fig:MF_Radial} and \tabref{tab:masssegregation}, the MF of low-mass stars at smaller radii is steeper compared to larger radii. In conjunction with the mean stellar mass which gets smaller at larger radii, these results are clear indications of mass segregation in Alpha Persei. Mass segregation is also evident in \figref{fig:Density} which shows that the surface density of bright stars $M>0.84\Msun$ sharply decreases as a function of radius compared to faint stars. 

\begin{figure}		
	\centering
	\includegraphics[width=0.55\textwidth]{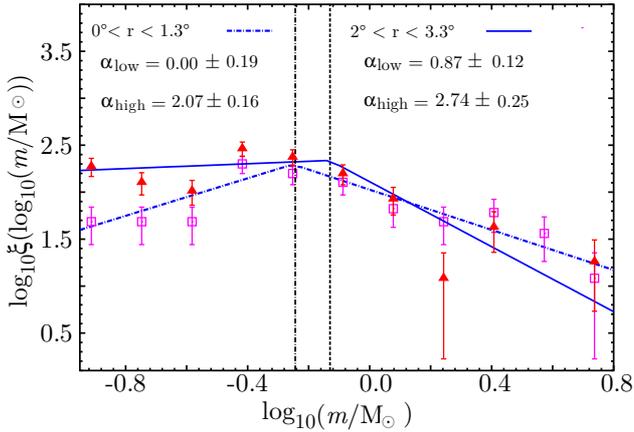}		
	\caption{The MF of Alpha Persei as a function of radius from the centre of the cluster. The variation of the MF slopes for low-mass and high-mass stars from inner to outer radii is the evidence of mass segregation.}
	\label{fig:MF_Radial}
\end{figure}

\begin{table}
	\caption{Effects of mass segregation on the average mass and the MF slopes of Alpha Persei as a function of radius from the cluster centre for two different binning schemes. Mass segregation is evident regardless of the binning scheme.}
	\label{tab:masssegregation}
	\centering
	\begin{tabular}{ccccccc}
		\hline $R$ & $N$ & $m_{\rm total}$ & $m_{\rm avg}$ & $\alphalow$ & $\alphahigh$ \\
  	($\deg$) & - & ($\Msun$) & ($\Msun$) & - & - \\
		\hline
    \textbf{I} &   &  &  &   &   \\
  	0.00-0.66 & 40 & 40.84 & 1.02 & $0.90\pm0.47$ & $2.61\pm0.37$ \\
  	0.66-1.32 & 105 & 94.14 & 0.90 & $0.00\pm0.22$ & $2.02\pm0.21$ \\
		1.32-1.98 & 103 & 73.50 & 0.71 & $0.31\pm0.20$ & $2.49\pm 0.32$ \\
  	1.98-2.64 & 107 & 73.02 & 0.68 & $0.70\pm0.19$ & $2.59\pm0.27$ \\
  	2.64-3.30 & 114 & 54.52 & 0.48 & $0.85\pm0.17$ & $2.52\pm0.46$ \\	\\
		\textbf{II} &   &  &  &   &   \\
    0.00-1.10 & 112 & 106.36 & 0.95 & $0.00\pm0.23$ & $2.01\pm0.21$ \\
    1.10-2.20 & 172 & 124.07 & 0.72 & $0.49\pm0.15$ & $2.59\pm0.21$ \\
    2.20-3.30 & 185 & 105.60 & 0.57 & $0.81\pm0.14$ & $2.63\pm 0.29$ \\    
    \hline
	\end{tabular}
\end{table}

\section{Binary Fraction}\label{sec:binary}			
\subsection{Unresolved binaries on CMD}
A large number of stars that belong to the galactic field and star clusters are found in binary systems \citep{Kroupa01, Kouwenhoven10}. These systems will influence the observed MF of star clusters \citep{Kroupa02}. As a result, it is crucial that the we estimate the effect of these systems on the measured MF of Alpha Persei to find the true underlying MF of the cluster.

\par In a CMD, binary systems populate a region between the single star main sequence and an isochrone which is $0.753\,\mathrm{mag}$ above. As it can be seen from \figref{fig:CMD1}, there are a number of stars, which are denoted by black triangles, whose vertical deviations from the isochrone are larger than the photometric errors and less than $0.753\,\mathrm{mag}$. Hence they can be identified as binaries. Defining the binary fraction as $f_{\rm bin} = N_{\rm bin}/(N_{\rm sin}+N_{\rm bin})$ we derive an observed binary fraction of 8 {\percent} for Alpha Persei in the mass range $0.7<\mratio<4.68$.

\par \citet{Hurley} created synthetic CMDs of binary systems with different mass ratios ($q$), i.e. the mass of the secondary component divided by that of the primary (most massive) component $q=m_{\rm prim}/m_{\rm sec}$. They concluded that a large fraction of their binaries, with faint secondary components, lie very close to the single star main sequence. Given that the magnitudes of stars in a CMD are subject to photometric errors, there could be many binaries whose vertical deviations from the single star main-sequence are less than their photometric errors. Hence they have been mistakenly identified as single stars in our analysis. Hereafter we refer to such systems as unresolved binaries. As a result the binary fraction of Alpha Persei can be larger than the observed binary fraction and the obtained MF slopes in \secref{sec:MF} need to be corrected for the effect of unresolved binaries.

\par We did a Monte Carlo analysis, explained in detail in \secref{sec:simulation}, to find true binary fractions and the underlying MF of Alpha Persei.

\begin{figure}	
	\includegraphics[width=0.45\textwidth]{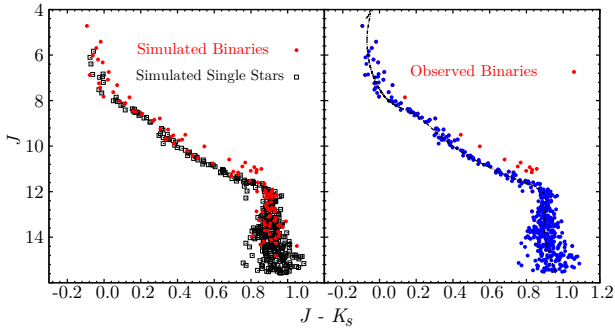}	
	\caption{(Left) Synthetic CMDs of simulated binaries (red filled circles) and single stars (black open squares) for a true binary fraction of $23$ {\percent} over the whole mass range $(0.1<\mratio<4.68)$. The scatter of single stars is due to the added photometric errors which resemble actual photometric errors in PPMXL. (Right) The detected binaries (red filled circles) as well as single stars (blue filled circles) after applying the blind photometric test on the simulated data . As it can be seen, many of the binaries have been identified as single stars. The observed binary fraction is $\sim6$ {\percent}. The PARSEC isochrone which is used in the blind photometric test, is shown by the black solid line.}
	\label{fig:simulation}
\end{figure}


\subsection{Monte Carlo simulation of binary fraction}\label{sec:simulation}
In this section, we explain the details of our Monte Carlo simulations to recover the underlying MF and binary fraction of the cluster.

\par We first make a sample of single star masses using a two-stage power-law MF (\equref{eq:powerlaw}) and convert these masses into $J$ and $K_s$ magnitudes using the PARSEC isochrones. We then pair the single stars to make binaries assuming a non-random pairing of the binary components. According to the hydrodynamical simulations of star cluster formation by \cite{Bate}, the multiplicity fraction is approximately proportional to the logarithm of the primary mass. As a result, we assume that the probability that a single star with mass $m_{\rm prim}$ is the primary component of a binary, scales linearly with the logarithm of its mass, i.e. ${\rm Prob}\propto \log_{10}(m_{\rm prim})$. We also assume a flat distribution function for the mass ratio $0\leq q\leq 1.0$. We then calculate the total $J$ and $K_s$ magnitudes of each binary from the sum of the luminosities of its components. We also add photometric errors to the magnitudes of single and binary stars. The photometric errors resemble the photometric errors of PPMXL at each magnitude. Using this procedure, we make a synthetic CMD of a cluster with single stars and binary systems. We then apply our photometric membership criterion (\secref{sec:photometric}) on this synthetic CMD through a blind test and count the number of single stars and binaries. Photometrically, a single star is any point that lies within $2.5\sigma$ of the isochrone and a binary is any non-single star whose vertical deviation from the isochrone is less that $0.75\,\mathrm{mag}$. Finally we change the low-mass and the high-mass slopes of the MF ($\alphalow$, $\alphahigh$) in ranges $(0.70,1.25)$ and $(2.00, 2.60)$ respectively, until we find the best match with the observed MF (using K-S test) and the observed binary fraction. For each set of the simulation parameters we have simulated 40 clusters with different random seed numbers, to reduce the errors caused by statistical fluctuations.

\par \figref{fig:simulation}, shows the synthetic CMD of one of our simulated clusters (left panel) and the detected single stars and binaries after applying the blind photometric test. In this figure, the true binary fraction is $23$ {\percent} in the mass range $0.10\leq M/\Msun\leq4.68$, but the observed binary fraction is only $\sim6$ {\percent}. 

\par \tabref{tab:binaryfraction} compares the binary fractions of Alpha Persei from different studies. According to the outcome of our Monte Carlo simulations, the binary fraction of the cluster over the mass-range $0.70<\mratio<4.68$ is $34\pm12$ {\percent}, in agreement with \citet{Mermilliod} results. This is comparable with $f_{\rm bin}=35$ {\percent} derived for Praesepe using the same method and over the same mass-range. Over the whole mass range $(0.10<\mratio<4.68)$, the binary fraction is $23\pm9$ {\percent}. In addition, we found that the unresolved binaries have a significant effect on the MF (see \tabref{tab:massfunction}). As one can see, this effect is especially important for the low-mass slope of the MF ($\alphalow$) which is in agreement with the simulations of \citetalias{Khalaj} (for Praesepe) and \citet{Zonoozi14} results (for Palomar 4). We find the total mass of Alpha Persei to be $336\Msun$. After considering the effect of unresolved binaries and correcting the MF for the effect that the binaries have on the inferred stellar masses, the cluster mass will be equal to $352\Msun$.

\par For comparison, we have repeated our Monte-Carlo analysis by combining the low-mass data of \citet{Lodieu12b} in the mass-range $0.1<\mratio<0.62$ with our high-mass data ($0.62<\mratio<4.68$) and determined the low-mass MF slope and binary fraction. We found that in this case, the low-mass MF slope after correction for binaries, becomes $\alphalow=0.75\pm12$ which is roughly in agreement with the value of $\alphalow=0.89\pm0.11$ that we derived for our low-mass data within 1-sigma. For the high-mass data, the MF slope changes to $2.39\pm0.05$, consistent with $2.37\pm0.09$. In addition, the binary fraction over the mass-range $0.70<\mratio<4.68$ and the whole mass-range become $36\pm12$ and $16\pm8$ {\percent} respectively, both of which are in agreement with the values we found earlier within 1-sigma. The total mass of the cluster in this case will be $463\Msun$.

\begin{table}
	\centering
 	\caption{The binary fraction of Alpha Persei from other studies}
 	\label{tab:binaryfraction}
	\begin{tabular}{ccc}
		\hline
		Reference & Binary fraction \\
		\hline		
        \citet{Morrell} & $14\%$ \\
		\citet{Makarov} & $20\%$ (for 28 brightest members) \\
		\citet{Martin} & $<9\%$ (brown dwarf binaries) \\
		\citet{Mermilliod} & $32\%$ ($9.5<V<13.85$) \\	
        Our work & $34\pm12\%$ $(0.70<\mratio<4.68)$ \\
        -- & $23\pm9\%$ $(0.10<\mratio<4.68)$ \\
		\hline
	\end{tabular}	
\end{table}

\begin{table}
	\centering
 	\caption{MF of Alpha Persei before and after considering the contribution of unresolved binaries.}
 	\label{tab:massfunction}
	\begin{tabular}{lll}				
		\hline
		mass range & $0.10\leq M/\Msun<0.62$ & $0.62\leq M/\Msun<4.68$ \\   
		slopes & $\alphalow=+0.5\pm0.09$ & $\alphahigh=+2.32\pm0.14$ \\ 		
		corrected slopes & $\alphalow=+0.89 \pm 0.11$ & $\alphahigh=+2.37\pm0.09$\\ 		
		\hline		
	\end{tabular}	
\end{table}

\section{Summary}\label{sec:summary}

Using the available proper motions from PPMXL catalogue and photometric data from WISE survey, the photometric and proper motion membership for the Alpha Persei is derived from the analysis of the PPMXL $(J, J-K_s)$ and WISE $(W1, W1-W2)$ CMDs. After applying astrometric and photometric membership criteria and subtracting the background stars, 810 possible member stars (down to $J=15.5$) are identified for Alpha Persei within its tidal radius of $3\fdg3$. 

\par We found evidence for a break of the MF at $m=0.62\Msun$. In addition, we found that the stellar MF of Alpha Persei is well fitted by a two-stage power law in the mass range $0.1-0.62\Msun$ and $0.62-4.68\Msun$ with $\alphalow=0.50\pm0.09$ and $\alphahigh=2.32\pm0.14$. This is significantly shallower than a \citet{Kroupa01} IMF in the low mass range with $\alphalow=2.35$. 

\par Moreover, deriving the stellar MF in different radial bins, we found evidence for mass segregation in this open cluster. The MF of low-mass stars at different radii is well approximated by a two-stage power-law and rises with radius from a shallow slope of $\alphalow=0.00\pm0.23$ in the region $r\leq1\fdg1$ to a slope of $\alphalow=0.81\pm0.14$ for $1\fdg1\leq r\leq2\fdg2$. The high-mass slope of the radial MF also rises from the inner to the outer radii, indicating the mass segregation in Alpha Persei.

\par Finally, a series of Monte Carlo simulations have been done to find the true binary fraction and the underlying MF of the cluster. The best-fitting model with observed MF shows that the fraction of binaries is $34\pm12$ {\percent} for $0.70<\mratio<4.68$ and and over the whole mass-range ($0.10<\mratio<4.68$) it is $23\pm9$ {\percent}.

\par By summing the masses of member stars and after considering the effect of unresolved binaries we estimated the total cluster mass (in the observed stellar mass range inside the projected tidal radius) of $352\Msun$. 

\par Having the present-day structural properties such as the global and radial MF and binary fraction for Alpha Persei and other well-studied open clusters such as Praesepe \citep{Boudreault, Khalaj} and Pleiades \citep{Lodieu12a, Bouy}, what remains to be seen is what this means for how the clusters started. Did all the clusters start from the same condition or were there variations? Did the clusters start mass segregated or are the present-day observations compatible with no primordial mass segregation? Determining the starting conditions (e.g. the initial MF-slopes, binary fractions, and primordial mass segregation at $T=0$) under which these clusters are formed and evolved by means of the {\Nbody} simulation is our upcoming project (Zonoozi et al., in preparation). According to \citet{Kroupa01}, the scatter introduced by Poisson noise and the dynamical evolution of star clusters produces quite well the observed scatter in the determination of the MF power-law index. As a result, the MF of open clusters is Salpeter-like down to $m\sim0.5-0.8\Msun$ and much flatter for low-mass stars. As an example, one can compare the MF of Alpha Per with Praesepe. The high-mass slope of Alpha Persei as found in this study is $\alphahigh=2.37\pm0.09$, whereas the high-mass slope of Praesepe is $\alphahigh=2.80\pm0.05$ \citepalias{Khalaj}. In addition, the low-mass slope of Praesepe ($\alphalow=1.05\pm0.05$; \citealt{Khalaj}) is flatter compared to Alpha Persei ($\alphalow=0.87\pm0.11$). These MF slopes are corrected for unresolved binaries using the procedure explained in \secref{sec:simulation}. Given the error-bars on the measured MFs, the difference between the MF of Alpha Persei and Praesepe is statistically significant, especially at the high-mass end. Since Praesepe is an older cluster ($T=590\Myr$; \citealt{Fossati}) compared to Alpha Persei ($T=100\Myr$; \citealt{Barrado02}), the observed difference in the measured MF can be attributed to stellar evolution, which makes the cluster devoid of high-mass stars as the cluster ages, hence making the high-mass slope steeper and low-mass slope flatter. A similar effect can be seen when one compares the mass function of Trapezium ($T=1.0\Myr$; \citealt{Muench}) and Pleiades ($T=120\Myr$; \citealt{Bouy}). Pleiades is an older cluster and has a steeper mass function at the high-mass end. 

\par Binaries can also play a role in the depletion of massive stars and making the high-mass slope steeper. Both Alpha Persei and Praesepe, have a binary fraction of $\sim30-35$ {\percent} ($m>0.7\Msun$) which can affect dynamical evolution of the clusters. Moreover, mass segregation is more evident in Praesepe. For Praesepe the high-mass slope changes from $\alphahigh=2.32\pm0.24$ inside the half-mass radius ($\sim1\fdg2$) to $\alphahigh=4.90\pm0.51$ outside the half-mass radius \citepalias{Khalaj}. This change in the high-mass slope of Praesepe is significantly larger than that of Alpha Persei (\tabref{tab:masssegregation}). This can be due to dynamical evolution or the way in which star formation has proceeded, i.e. primordial mass segregation. {\Nbody} simulations are therefore needed to further elaborate on this.

\section*{Acknowledgements}
We would like to thank the anonymous referee for his/her useful comments and suggestions which improved the quality of this work.


\bibliographystyle{mnras}
\input{manuscript.bbl}


\bsp
\label{lastpage}

\end{document}